\documentclass[conference]{IEEEtran}
\IEEEoverridecommandlockouts

\usepackage[numbers,sort&compress,square]{natbib}
\usepackage{amsmath,amssymb,amsfonts}
\usepackage{algorithmic}
\usepackage{balance}
\usepackage{booktabs}
\usepackage[hidelinks]{hyperref}
\usepackage{tikz}
\usetikzlibrary{positioning}
\usepackage[ruled,vlined,linesnumbered]{algorithm2e}
\usepackage{graphicx}
\usepackage{subcaption}
\usepackage{graphicx}
\usepackage{textcomp}
\usepackage{xcolor}
\usepackage[authormarkup=none]{changes}
\definechangesauthor[name={Jyoti}, color=blue]{JS}

\def\BibTeX{{\rm B\kern-.05em{\sc i\kern-.025em b}\kern-.08em
  T\kern-.1667em\lower.7ex\hbox{E}\kern-.125emX}}
 \usepackage{xcolor}
\usepackage[authormarkup=none]{changes}

\definecolor{mygreen}{RGB}{90,180,120}

\definechangesauthor[name={JYX}, color=mygreen]{YX}
\usepackage{bbding} 
\usepackage[misc]{ifsym}
\newcommand{\corr}{$^\text{\Envelope}$}

\begin{document}

\title{Hierarchical Intelligent Evolutionary Optimization for UAV-Assisted NOMA-MEC Systems\\
}
\author{
\IEEEauthorblockN{
Yaxuan Jiao\IEEEauthorrefmark{1},
Jiaxuan Lv\IEEEauthorrefmark{1},
Jyoti Sahni\IEEEauthorrefmark{2},
Wuyungerile Li\IEEEauthorrefmark{1}\corr\thanks{Corresponding author: gerile@imu.edu.cn},
Winston K.G. Seah\IEEEauthorrefmark{2}
}
\IEEEauthorblockA{\IEEEauthorrefmark{1}School of Computer Science, Inner Mongolia University, Hohhot, China\\
Email: 32409126@mail.imu.edu.cn, 1518371393@qq.com, gerile@imu.edu.cn}
\IEEEauthorblockA{\IEEEauthorrefmark{2}School of Engineering and Computer Science, Victoria University of Wellington, Wellington, New Zealand\\
Email: jyoti.sahni@vuw.ac.nz, winston.seah@ecs.vuw.ac.nz}
}
\maketitle

\begin{abstract}

The integration of unmanned aerial vehicles (UAVs), mobile edge computing (MEC) and non-orthogonal multiple access (NOMA) provides an effective solution for computation-intensive and latency-sensitive services {in infrastructure-deficient environments}. 
However, the joint optimization of UAV trajectories, task offloading, communication and computing resources {under heterogeneous task priorities} results in a highly coupled mixed-integer non-convex problem. 
%
To address these challenges, this paper proposes a Hierarchical Intelligent Evolutionary Algorithm (HIEA) for priority-aware multi-UAV cooperative NOMA-MEC networks{, which} integrates {multi-head deep Q-network (MH-DQN)-based }intelligent decision generation, communication-aware initialization, adaptive evolutionary exploration and  { successive convex approximation (SCA)-based }precision refinement. 
Specifically, a novel communication-aware initialization mechanism, termed {Channel Quality Indicator (}CQI)-guided Load-Aware Trajectory Initialization (CQI-LATI), is designed to incorporate CQI and task workload characteristics into evolutionary population construction, generating high-quality and diversity-preserving initial populations.
Theoretical analysis  {characterizes HIEA's search-space reduction, the quality-diversity trade-off of CQI-guided initialization, and the accuracy-efficiency trade-off of SCA refinement, }and extensive simulations demonstrate that HIEA outperforms benchmark schemes, with gains of up to 6.5\% in  throughput, 13.6\% in task completion rate and 15.3\% in energy efficiency.

\end{abstract}

\section{Introduction}

The rapid development of Internet of Things (IoT), intelligent transportation and emergency communication services has generated increasing demands for computation-intensive and latency-sensitive applications. {In such scenarios, tasks exhibit heterogeneous priorities, as emergency and mission-critical services demand stricter latency and reliability guarantees than routine IoT traffic.}
However, traditional cloud computing architectures suffer from long transmission latency and limited service availability, especially in remote and infrastructure-deficient environments. 
Mobile edge computing (MEC) enables low-latency services by deploying computing resources closer to end users~\cite{mao2017survey}. 
Meanwhile, unmanned aerial vehicles (UAVs), with flexible deployment capability and three-dimensional mobility, have emerged as promising platforms for providing communication and computing services in dynamic scenarios~\cite{zeng2016wireless}.

To support massive connectivity and improve spectrum utilization, non-orthogonal multiple access (NOMA) has been widely adopted in UAV-assisted MEC networks~\cite{chen2025qos,islam2018resource}. 
Consequently, the joint optimization of UAV trajectories, task offloading and computing resource management has become a fundamental problem in multi-UAV cooperative MEC systems~\cite{hao2024joint,pervez2023energy}. 
However, the strong coupling among heterogeneous decision variables makes this problem a large-scale mixed-integer non-convex optimization problem with a highly complex search space.

Existing studies have explored {decomposition-based,} heuristic, evolutionary and learning-based approaches for UAV-MEC optimization. {Decomposition-based methods, such as block coordinate descent with convex relaxation, decouple the problem into tractable subproblems~\cite{zhao2025joint}, but converge only to locally optimal solutions and are sensitive to initialization conditions. Heuristic methods, such as greedy and matching-based strategies, offer low complexity, yet provide no optimality guarantees and generalize poorly across network conditions.}
Evolutionary methods {offer} robust {global search for}  nonlinear {mixed-integer} optimization problems~\cite{zhu2021nsga,kuang2023utility,ning2021dynamic} but suffer from high computational cost and poor efficiency. 
Meanwhile, learning-based methods enable adaptive decision-making under dynamic network conditions by exploiting historical interactions~\cite{zhang2020three,zhong2021multi,huang2024multiobjective}, although end-to-end learning requires extensive training and struggles to guarantee constraint satisfaction.

{Evolutionary search thus best fits this problem class, and its limitations can be alleviated by incorporating prior knowledge~\cite{chen2023subspace} and learning techniques into the search framework.}
 {Realizing such an integrated framework, however, involves several challenges.} 
First, evolutionary performance depends heavily on population initialization and diversity preservation, which govern initial solution quality, convergence efficiency and the exploration-exploitation balance~\cite{tharwat2021population,elsayed2016sequence,lengler2023analysing}. 
Although knowledge-assisted initialization has demonstrated the value of prior information in population construction~\cite{chen2023subspace}, such approaches mainly exploit mathematical properties or historical optimization information, while rarely considering domain-specific knowledge such as wireless channel conditions, user distribution and task heterogeneity.

Second, existing UAV-MEC optimization studies mainly focus on optimizing final decision variables, such as UAV trajectories, offloading strategies and resource allocation, leaving the impact of communication-aware initialization on search efficiency largely unexplored. 
Random or geometry-based initialization may generate low-quality candidates, resulting in inefficient exploration in large-scale search spaces. 
Third, although local refinement techniques can improve the accuracy of the solution, they usually introduce additional computational complexity and limit scalability.

To address these challenges, this paper proposes a Hierarchical Intelligent Evolutionary Algorithm (HIEA) for priority-aware multi-UAV cooperative NOMA-MEC networks. 
Unlike conventional frameworks that mainly improve final solutions, HIEA enhances the whole evolutionary process through four coordinated functions{: multi-head deep Q-network (MH-DQN)-based intelligent decision generation, communication-aware initialization, adaptive evolutionary exploration and successive convex approximation (SCA)-based precision refinement.
}

The key innovation of HIEA lies in the proposed communication-aware initialization mechanism. 
By jointly exploiting wireless channel quality and task workload characteristics, the proposed mechanism constructs feasible and diverse initial populations, {introducing communication-oriented guidance into population construction and adaptively balancing initialization quality and population diversity. Building on this initialization, the remaining functions jointly drive the search: MH-DQN injects learned decision candidates into the population, adaptive evolution performs global exploration, and SCA-based refinement sharpens elite solutions.}

The main contributions of this paper are summarized as follows:
\begin{itemize}

\item We formulate a priority-aware joint optimization problem for multi-UAV cooperative NOMA-MEC networks, jointly optimizing UAV trajectories, task offloading and computing resource management.

\item {We propose HIEA, a hierarchical framework that coordinates MH-DQN-based intelligent decision generation, communication-aware initialization, adaptive evolutionary exploration and SCA-based precision refinement to solve the formulated mixed-integer non-convex problem.}

\item Within HIEA, we design a new communication-aware initialization mechanism, termed Channel Quality Indicator (CQI)-guided Load-Aware Trajectory Initialization (CQI-LATI), which incorporates CQI and task workload characteristics into evolutionary population construction.

\end{itemize}

{Theoretical analysis evaluates MH-DQN search-space reduction, CQI-guided initialization, and SCA refinement trade-offs.
This is further supplemented by extensive simulations that demonstrate} 
{HIEA outperforms benchmark schemes in throughput, task completion rate and energy efficiency.}

\section{SYSTEM MODEL AND PROBLEM FORMULATION}

We consider a NOMA-enabled multi-UAV mobile edge computing (UAV-MEC) system consisting of a set of ground users $\mathcal{K}=\{1,\ldots,K\}$ and a set of UAVs $\mathcal{U}=\{1,\ldots,U\}$. Due to the lack of terrestrial infrastructure, users rely on UAVs for task offloading and computation services. The UAVs dynamically adjust their three-dimensional (3D) positions to provide communication and computing support.

Users are partitioned into $U$ NOMA clusters according to UAV association. Let $\mathcal{C}_u$ denote the user cluster served by UAV $u$. Each user is associated with only one UAV and each UAV serves one cluster. All UAVs operate on the same frequency band and thus inter-UAV interference is considered.

The system operates in discrete time slots $\mathcal{T}=\{1,\ldots,T\}$. Users move randomly within the service area and generate computation tasks with different priorities. The task generated by user $k$ at time slot $t$ is represented as
$q_k(t) = \big( c_k(t),\, n_k(t),\, m_k(t),\, o_k(t) \big)$
where $c_k(t)$, $n_k(t)$, $m_k(t)$ and $o_k(t)$ denote the required computation (CPU cycles), input data size, latency requirement and task priority, respectively. Since mobile devices have limited computing capability, tasks can be offloaded to UAVs for execution.

The air-to-ground (A2G) channel between users and UAVs follows a probabilistic path-loss model incorporating both line-of-sight (LoS) and non-line-of-sight (NLoS) links~\cite{zhang2020three}.

At time slot $t$, let $(x_u(t),y_u(t),Z_u(t))$ denote the 
three-dimensional position of UAV $u$, where $x_u(t)$ and $y_u(t)$ represent the horizontal coordinates and $Z_u(t)$ represents the flight altitude.  The horizontal distance between MU $k$ and UAV $u$ is denoted by
$d_{k,u}(t)=\sqrt{(x_u(t)-x_k(t))^2+(y_u(t)-y_k(t))^2}$.
The elevation angle is defined as
$\kappa_{k,u}(t)=\frac{180}{\pi}\arctan(\frac{Z_u(t)}{d_{k,u}(t)})$.
Then, the LoS probability between MU $k$ and UAV $u$ at time slot $t$ is
given by\vspace{-1ex}
\begin{equation}
P_{k,u}^{\mathrm{LoS}}(t)=
\begin{cases}
1, & d_{k,u}(t)\leq d_0,\\
\frac{1}{1+a\exp\big(-b(\kappa_{k,u}(t)-a)\big)},
& d_{k,u}(t)>d_0,
\end{cases}
\end{equation}
where $a$ and $b$ are environment-dependent constants and $d_0$ is the horizontal distance threshold for guaranteed LoS transmission.

Accordingly,
the NLoS probability is given by
$P_{k,u}^{\mathrm{NLoS}}(t)=1-P_{k,u}^{\mathrm{LoS}}(t)$.
Following the air-to-ground channel model in~\cite{zhong2021multi}, the path loss between MU $k$ and UAV $u$ under LoS and NLoS conditions (in dB) are respectively given by
\begin{align}
L^{\mathrm{LoS}}(t)
&= 30.9 + \left(22.25 - 0.5 \log_{10} Z_u(t)\right)\log_{10} d \nonumber \\
&\quad + 20 \log_{10} f_c,
\end{align}
\begin{align}
L^{\mathrm{NLoS}}(t)
&= \max \Big\{ L^{\mathrm{LoS}}(t), \nonumber \\
&\quad 32.4 + \left(43.2 - 3.1 \log_{10} Z_u(t)\right)\log_{10} d \nonumber \\
&\quad + 20 \log_{10} f_c \Big\},
\end{align}
where $d$ denotes the 3D distance between MU $k$ and UAV $u$, given by
$d = \sqrt{d_{k,u}^2(t) + Z_u^2(t)}$, and $f_c$ is the carrier frequency.
Therefore, the average path loss is expressed as
\begin{equation}
PL_{k,u}(t) = P_{k,u}^{\mathrm{LoS}}(t)\, L^{\mathrm{LoS}}(t)
+ P_{k,u}^{\mathrm{NLoS}}(t)\, L^{\mathrm{NLoS}}(t).
\end{equation}

Considering small-scale fading, the channel gain between MU $k$ and UAV $u$ at time slot $t$ is given by
\begin{equation}
g_{k,u}(t) = H_{k,u}(t) \cdot 10^{-\frac{PL_{k,u}(t)}{10}},
\end{equation}
where $H_{k,u}(t)$ denotes the small-scale fading power gain between UAV $u$ and MU $k$~\cite{zhong2021multi}.

\subsection{MU and UAV Mobility Models}
{UAV and MU GPS locations 
are used for cluster association.} Users are grouped into UAV-associated clusters according to their spatial distribution. {Both random-walk and directional mobility models are considered, following the mobility models in~\cite{chen2025qos}.} 
For the random mobility model, the movement direction and speed of MU $k$ at each discrete time slot $t$ are randomly generated, following uniform distributions  
{$\theta_k(t) \sim \mathcal{U}(0, 2\pi)$ and $V_k^{\mathrm{MU}}(t) \sim \mathcal{U}(0, V_{\max}^{\mathrm{MU}})$, respectively.}
{For the directional mobility model, the movement of each user at time slot
$t$ follows a vectorial mobility model, where the velocity is determined
by the combination of a dominant direction component and a random
perturbation component. Specifically, the velocity vector is expressed as
$\mathbf{V}_k(t)=\mathbf{D}_k(t)\!+\!\mathbf{\epsilon}_k(t)$, where
$\mathbf{D}_k(t)$ and $\mathbf{\epsilon}_k(t)$ denote the dominant direction vector
and random perturbation vector, respectively. The dominant direction
vector $\mathbf{D}_k(t)$ is introduced to represent the main moving
tendency of MU $k$. 
Notably, the direction and magnitude satisfy
$\theta_k^{D}\!(t)\!=\!\Theta$ and
$\|\mathbf{D}_k(t)\|\!=\!\frac{4}{5}V_{\max}^{\mathrm{MU}}$,
where $\Theta$ denotes the predefined dominant moving direction of MU $k$.
The random perturbation vector $\mathbf{\epsilon}_k(t)$ has a random direction
$\theta_k^{\epsilon}(t)\sim\mathcal{U}(0,2\pi)$ and a magnitude
$\|\mathbf{\epsilon}_k(t)\|\sim\mathcal{U}(0,\frac{1}{5}V_{\max}^{\mathrm{MU}})$.
The position of MU $k$ at time slot $t$ is denoted as
$\mathbf{M}_k(t)=[x_k(t),y_k(t),0]$.
Let $d_k(t)$ denote the traveling distance of MU $k$ during time slot
$t$. Then, the position update at time slot $t+1$ is given by}
\begin{align}
x_k(t+1) &= x_k(t) + d_k(t)\cos\big(\theta_k(t)\big) \quad \forall k,t \\
y_k(t+1) &= y_k(t) + d_k(t)\sin\big(\theta_k(t)\big) \quad \forall k,t
\end{align}

At time slot $t$, the three-dimensional position of UAV $u$ is denoted as
$\mathbf{q}_u(t)=(x_u(t),y_u(t),z_u(t))$.
Different from mobile users with stochastic mobility patterns, UAVs act as
controllable aerial MEC servers and their trajectories are optimized to
improve communication and computation performance.
The UAV trajectory evolution is expressed as
\begin{equation}
\mathbf{q}_u(t+1)
=
\mathbf{q}_u(t)+\mathbf{v}_u(t)\Delta t,
\quad \forall u,t,
\end{equation}
where $\mathbf{v}_u(t)$ denotes the velocity vector of UAV $u$ and
$\Delta t$ is the duration of one time slot.

To ensure that UAVs remain within the rectangular service area, the horizontal coordinates must satisfy
\begin{equation}
0 \leq x_u(t) \leq X_{\max}, \quad 0 \leq y_u(t) \leq Y_{\max} \label{eq:svc_area}
\end{equation}

Moreover, the altitude of UAVs is restricted within the allowable flight
range:
\begin{equation}
z_{\min}\leq z_u(t)\leq z_{\max},
\quad \forall u,t .
\label{eq:svc_areaz}
\end{equation}

To guarantee feasible UAV movement, the displacement between two
consecutive time slots is constrained by the maximum flight speed:
\begin{equation}
\left\|
\mathbf q_u(t+1)-\mathbf q_u(t)
\right\|
\leq V_{\max}^{\mathrm{UAV}}\Delta t,
\quad \forall u,t .
\label{eq:uav_speed}
\end{equation}
where $V_{\max}^{\mathrm{UAV}}$ denotes the maximum flight speed of UAVs.

Furthermore, to avoid collisions among UAVs, the distance between any two
UAVs should be larger than a predefined safety distance:
\begin{equation}
\left\|
\mathbf q_m(t)-\mathbf q_n(t)
\right\|
\geq D_{\min},
\quad \forall m\neq n,t .
\label{eq:uav_separation}
\end{equation}


\subsection{NOMA Transmission Model}

MUs access UAVs via uplink NOMA. Following the SIC
principle, each UAV decodes users in descending channel-gain order and
successively removes the signals of already decoded users.

Consider users associated with UAV $u$, indexed according to their channel gains as
\begin{equation}
{g_{1,u}(t) \geq g_{2,u}(t) \geq \cdots \geq g_{K,u}(t).\label{eq:noma_sic_order}}
\end{equation}
The SINR of user $k$ at time slot $t$ is
\begin{equation}
\gamma_{k,u}(t) = 
\frac{g_{k,u}(t)\, p_k(t)}
{I_{k,u}^{\mathrm{intra}}(t) + I_{k,u}^{\mathrm{inter}}(t) + \sigma^2},
\end{equation}
where $p_k(t)$ is the transmit power, $\sigma^2$ is the AWGN power,
$I_{k,u}^{\mathrm{intra}}(t) = \sum_{j = k+1}^{K} g_{j,u}(t)\, p_j(t)$ is intra-UAV interference from weaker users,
and
{{$I_{k,u}^{\mathrm{inter}}(t) = \sum_{s \neq u} \sum_{i \in \mathcal{C}_s} g_{i,u}(t)\, p_i(t)$ }}is inter-UAV interference.
The uplink rate is given by
\begin{equation}
R_{k,u}(t) = B \log_2 \big( 1 + \gamma_{k,u}(t) \big),
\end{equation}
where $B$ is the system bandwidth.

\subsection{User Association and Computation-Energy Model}

A binary variable $A_{k,u}(t)\in\{0,1\}$ 
indicates whether user $k$ is associated with UAV $u$ at slot $t$, subject to
\begin{equation}
\sum_{u=1}^{U} A_{k,u}(t) = 1, \quad \forall k,t.\label{eq:user_association}
\end{equation}
The user set of UAV $u$ is $\mathcal{C}_u(t)=\{k \mid A_{k,u}(t)=1\}$.
Let $W_{k,u}(t)\in\{0,1\}$ denote offloading from user $k$ to UAV $u$, with feasibility constraint $W_{k,u}(t)\le A_{k,u}(t)$.
Define $W_k(t)=\sum_u W_{k,u}(t)\in\{0,1\}$, then the delay of user $k$ at slot $t$ is given by
\begin{equation}
\begin{aligned}
T_k(t) =& (1-W_k(t))\frac{c_k(t)}{f_k^{\mathrm{loc}}} \\
&+ \sum_{u=1}^{U} W_{k,u}(t)\left(\frac{n_k(t)}{R_{k,u}(t)}+\frac{c_k(t)}{f_{k,u}^{\mathrm{UAV}}(t)}\right),
\end{aligned}
\end{equation}
where $c_k(t)$ and $n_k(t)$ are the required CPU cycles and input data size, respectively.

The user energy is
\begin{equation}
\begin{aligned}
E_k(t)=&
(1-W_k(t))\kappa_{\mathrm{m}}
\big(f_k^{\mathrm{loc}}\big)^2 c_k(t)\\
&+\sum_{u=1}^{U} W_{k,u}(t)
\big(P_k^{\mathrm{tx}}(t)+P_k^{\mathrm{proc}}(t)\big)
T_{k,u}^{\mathrm{tx}}(t),
\end{aligned}
\end{equation}
where $T_{k,u}^{\mathrm{tx}}(t)=n_k(t)/R_{k,u}(t)$.

For UAV $u$, the flight power is
\begin{equation}
P_u^{\mathrm{fly}}(t)=
\begin{cases}
P_{\mathrm{hover}}+\alpha V_u^2(t)+\beta V_u^3(t), & V_u(t)>0,\\
P_{\mathrm{hover}}, & \text{hovering},
\end{cases}
\end{equation}
and the slot energy is
\begin{equation}
E_u(t)=P_u^{\mathrm{fly}}(t)\Delta t
+ \sum_{k} W_{k,u}(t)\kappa_{\mathrm{u}}\big(f_{k,u}^{\mathrm{UAV}}\big)^2 c_k(t).
\end{equation}

Therefore, the system throughput and total energy are
\begin{equation}
R_{\mathrm{sum}}=\sum_{t=1}^{T}\sum_{k\in\mathcal{K}} R_k(t),
\end{equation}
\begin{equation}
E_{\mathrm{tot}}=\sum_{t=1}^{T}\left(\sum_{u}E_u(t)+\sum_{k}E_k(t)\right).
\end{equation}

\subsection{Priority-Aware Satisfaction Model}
The satisfaction model is adapted from the delay-sensitive utility model
in~\cite{hao2024joint} and further extended to capture heterogeneous task
priorities by introducing distinct satisfaction functions for low- and
high-priority tasks.

To characterize heterogeneous task priorities, let 
$o_k(t)\in\{0,1\}$ denote the priority indicator, where 
$o_k(t)=1$ represents a high-priority task.
Define the delay deviation as
\[
\eta_k(t)=T_k(t)-m_k(t),
\]
where $m_k(t)$ is the maximum tolerable delay.
The user satisfaction is modeled as

\begin{equation}
\Phi_k(t)
=
(1-o_k(t))\Phi_k^{\mathrm{L}}(t)
+
o_k(t)\Phi_k^{\mathrm{H}}(t),
\end{equation}

where

\begin{equation}
\Phi_k^{\mathrm{L}}(t)=
\begin{cases}
P_L, & \eta_k(t)\leq0,\\
P_L e^{-\rho\eta_k(t)}, & \eta_k(t)>0,
\end{cases}
\end{equation}

\begin{equation}
\Phi_k^{\mathrm{H}}(t)=
\begin{cases}
P_L+P_H\log_2(1-\eta_k(t)),
& \eta_k(t)\leq0,\\
-P_H,& \eta_k(t)>0 .
\end{cases}
\end{equation}

The system-level satisfaction is given by

\begin{equation}
\Phi_{\mathrm{sum}}
=
\sum_{t=1}^{T}\sum_{k=1}^{K}\Phi_k(t).
\end{equation}

\subsection{Problem Formulation}
To maximize the overall system utility while reducing energy consumption, we jointly optimize the task offloading decisions, UAV three-dimensional trajectories, and computational resource allocation. The user--UAV association is generated by 
{a K-means-based user clustering mechanism} and dynamically updated via periodic re-clustering according to user mobility, as inspired by~\cite{zhong2021multi}.
\begin{subequations}\label{P1}
\begin{align}
\max_{\mathbf{W},\mathbf{Q},\mathbf{F}}
\quad
&
\omega_1
\frac{R_{\mathrm{sum}}}{R_{\mathrm{norm}}}
+
\omega_2
\frac{\Phi_{\mathrm{sum}}}{\Phi_{\mathrm{norm}}}
-
\omega_3
\frac{E_{\mathrm{tot}}}{E_{\mathrm{norm}}}
\notag
\\
\text{s.t.}\quad
&
\sum_{k=1}^{K} f_{k,u}^{\mathrm{UAV}}(t)
\leq f_u^{\max},
\quad \forall u,t,
\label{P1a}
\\
&
\eqref{eq:svc_area}-\eqref{eq:noma_sic_order},
\eqref{eq:user_association}.
\label{P1c}
\end{align}
\end{subequations}
Constraint \eqref{P1a} ensures that the computational resources allocated 
to users do not exceed the maximum CPU capacity of each UAV. 
Constraints \eqref{eq:svc_area}, \eqref{eq:uav_speed}, and 
\eqref{eq:uav_separation} guarantee the UAV mobility feasibility, 
including the horizontal boundary limitation, maximum flight speed 
constraint, and minimum safety distance between UAVs, respectively. 
Constraint \eqref{eq:svc_areaz} restricts the UAV flight altitude. 
Constraint \eqref{eq:noma_sic_order} ensures the correctness of the SIC 
decoding order in NOMA transmission. 
Constraint \eqref{eq:user_association} describes the user--UAV 
association and task offloading decisions.
\section{Hierarchical Intelligent Evolutionary Algorithm}
\label{sec:hiea}

\begin{figure}[b]
\centering    
\includegraphics[width=\columnwidth]{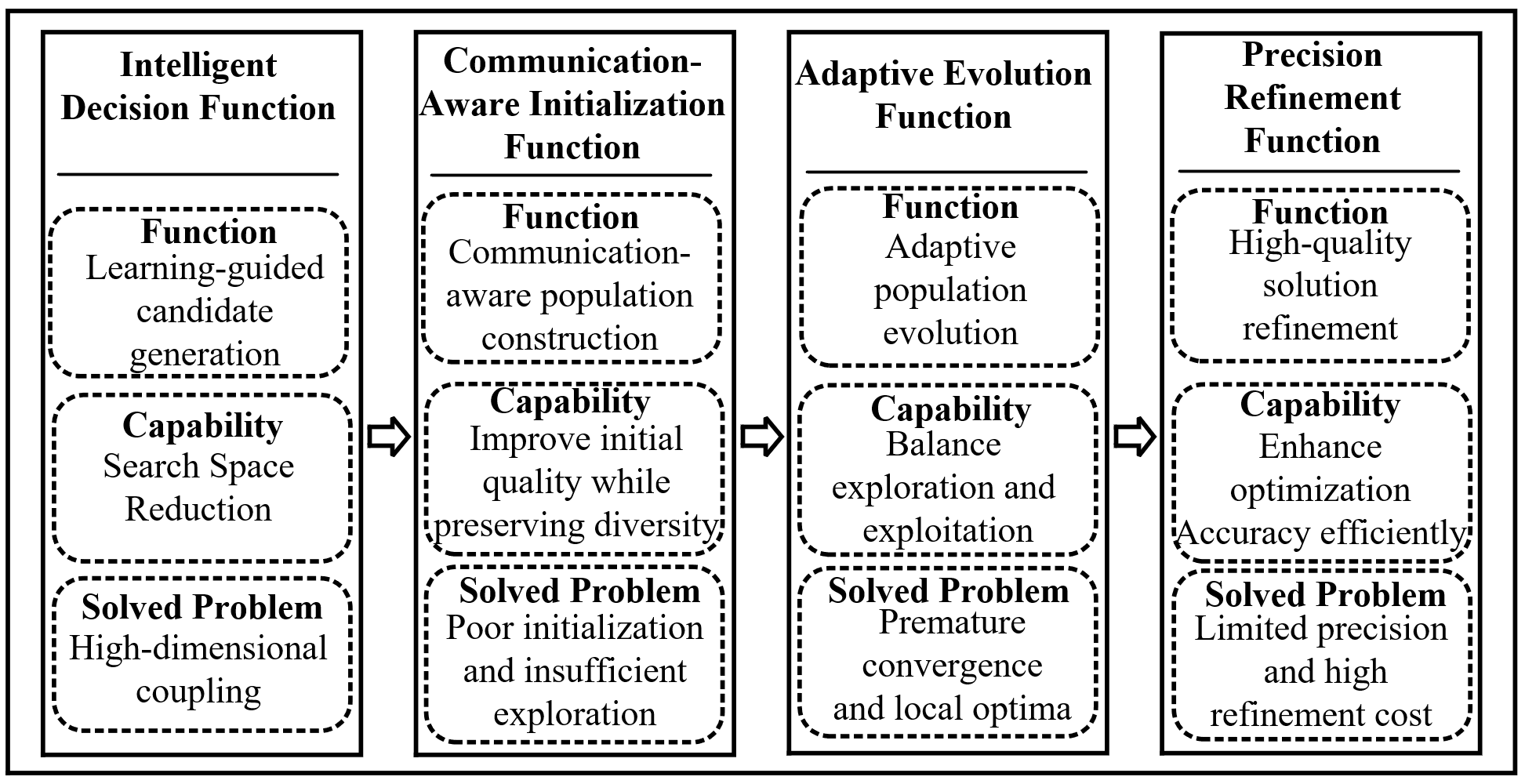}    
\caption{Functional architecture of HIEA}    
\label{fig:hiea_function}
\end{figure}

Fig.~\ref{fig:hiea_function} illustrates the architecture of the proposed HIEA framework. 
HIEA decomposes the optimization process into four coordinated functions, including intelligent decision, communication-aware initialization, adaptive evolution and precision refinement, to efficiently solve the mixed-integer non-convex optimization problem.

\subsection{Intelligent Decision Function}
\label{sec:idf}

The intelligent decision function aims to reduce the high-dimensional search complexity caused by the joint optimization of task offloading and UAV mobility. 
A conventional DQN can learn effective policies for sequential decision-making; however, directly applying DQN to the coupled offloading and mobility optimization results in a large mixed action space and severe decision interference. 
To address this issue, we propose a Multi-Head Deep Q-Network (MH-DQN) which decomposes heterogeneous decisions into multiple coordinated heads. 
Instead of directly solving the entire optimization problem, MH-DQN serves as a candidate generator to learn promising offloading and trajectory decisions and provide high-quality initial solutions for subsequent evolutionary optimization.

\subsubsection{State and action design}
The state consists of normalized UAV/user locations, user--UAV association, task features 
{{$(c_k,n_k,m_k,o_k)$}} and user identifiers. 
Different from conventional DQN with a single action output, MH-DQN adopts two decoupled decision heads: an offloading head with binary actions and a mobility head with seven movement actions. 
This design reduces action coupling and enables efficient learning of heterogeneous decisions.

\subsubsection{Network and reward design}
MH-DQN employs a shared feature extractor followed by two Q-value heads to capture the correlations and differences between offloading and mobility decisions. 
It is trained using experience replay and a target network. 
The reward is aligned with the global optimization objective:
\begin{equation}
J=
\omega_1\tilde{R}_{\mathrm{sum}}
+
\omega_2\tilde{\Phi}_{\mathrm{sum}}
-
\omega_3\tilde{E}_{\mathrm{tot}}
-\Pi,
\end{equation}
where $\tilde{R}_{\mathrm{sum}}$, $\tilde{\Phi}_{\mathrm{sum}}$, and 
$\tilde{E}_{\mathrm{tot}}$ denote the normalized throughput, 
priority-aware satisfaction, and energy consumption, respectively.
The penalty term $\Pi$ is used to handle constraint violations.

\subsubsection{Candidate generation}
After training, MH-DQN generates trajectory--offloading candidates over multiple time slots. 
These candidates are projected into feasible solutions and injected into the initial population, providing effective search guidance and reducing the optimization burden of the evolutionary process.

\begin{algorithm}[t]
\caption{CQI-Guided Load-Aware Trajectory Initialization (CQI-LATI)}
\label{alg:cqi_lati_en}
\begin{algorithmic}[1]

\REQUIRE User association $\mathbf A$, task features 
$\{(c_k,n_k,m_k,o_k)\}$, population size $N_p$
\ENSURE Initial population $\{\mathcal X_i\}_{i=1}^{N_p}$

\STATE Generate candidate service points and select CQI-optimal locations:
\[
\mathbf q_u^{\mathrm{CQI}}
=\arg\max_{\mathbf s\in\mathcal S_u}R_u(\mathbf s)
\]

\STATE Compute load-adaptive guidance factor $\alpha_u$

\FOR{$i=1$ to $N_p$}

    \STATE Generate UAV trajectories using fused random and CQI guidance:
    \[
    \Delta\mathbf q_u(t)
    =(1-\alpha_u)\Delta\mathbf q_{\rm rand}
    +\alpha_u\Delta\mathbf q_{\rm CQI}
    \]

    \STATE Project trajectories onto UAV mobility constraints

    \STATE Initialize $\mathbf W_i$ and $\mathbf f_i$, and construct
    $\mathcal X_i=(\mathbf Q_i,\mathbf W_i,\mathbf f_i)$

\ENDFOR

\RETURN $\{\mathcal X_i\}_{i=1}^{N_p}$

\end{algorithmic}
\end{algorithm}

\subsection{Communication-Aware Initialization Function}
\label{sec:cai}

The communication-aware initialization function aims to improve the quality of evolutionary populations by incorporating wireless channel conditions into UAV trajectory initialization.
Different from conventional random or geometry-based initialization, we propose a CQI-guided load-aware trajectory initialization (CQI-LATI) mechanism, which jointly considers communication quality, task heterogeneity and UAV mobility constraints.
The overall procedure is summarized in Algorithm~\ref{alg:cqi_lati_en}.

\subsubsection{Communication-aware service point selection}
Instead of selecting UAV locations merely according to user distribution, CQI-LATI identifies communication-efficient service points by evaluating the achievable NOMA transmission performance.
For each UAV-associated cluster, a candidate service-point set is generated around the user distribution center:
\begin{equation}
\mathcal{S}_u=
\{\mathbf s_{u,1},\mathbf s_{u,2},...,\mathbf s_{u,M}\},
\end{equation}
%
where each candidate point is generated as
$\mathbf s_{u,m}=(\bar{x}_u+\Delta x_m,\bar{y}_u+\Delta y_m,H_u)$.
The CQI-guided service point is selected by maximizing the aggregate achievable rate:

\begin{equation}
\mathbf q_u^{CQI}
=
\arg\max_{\mathbf s_{u,m}\in\mathcal S_u}
\sum_{k\in\mathcal C_u(t)}R_{k,u}(\mathbf s_{u,m}),
\end{equation}
where $R_{k,u}(\mathbf s_{u,m})$ denotes the achievable transmission rate of user $k$ at candidate location $\mathbf s_{u,m}$.
This enables the initialization process to exploit channel-aware spatial information rather than only geographical proximity.

\subsubsection{Load-aware trajectory guidance}
To adapt the trajectory exploration to heterogeneous task requirements, CQI-LATI introduces a priority-aware load coefficient:
\begin{equation}
 \alpha_u=
\frac{\sum_{k\in\mathcal C_u(t)}(1+\eta o_k)c_k}
{\max_j\sum_{k\in\mathcal C_j(t)}(1+\eta o_k)c_k},
\end{equation}
where $o_k$ represents the task priority indicator.
A larger $\alpha_u$ indicates heavier workload, leading to stronger CQI-oriented guidance, while lightly loaded UAVs preserve exploration diversity.
Therefore, CQI-LATI achieves an adaptive balance between communication-aware exploitation and population diversity.

\subsubsection{Velocity-constrained individual construction}
To avoid infeasible trajectories caused by direct position sampling,CQI-LATI adopts trajectory-increment encoding satisfying the UAV mobility constraint in \eqref{eq:uav_speed}.

After trajectory generation, the remaining decision variables,
including task offloading decisions and computational resource allocation,are initialized by the corresponding modules. The user--UAV association is inherited from 
the K-means-based user clustering results~\cite{zhong2021multi}. 
Finally, a feasible individual is obtained as
\begin{equation}
\mathcal X_i=(\mathbf Q_i,\mathbf W_i,\mathbf f_i).
\end{equation}
%
By integrating communication-aware spatial guidance and workload-adaptive exploration, CQI-LATI provides informative trajectory priors for HIEA,thereby improving the quality, diversity, and convergence efficiency of the initial population.




\subsection{Adaptive Evolution Function}
\label{sec:aef}

After initialization, HIEA employs an adaptive evolution function to enhance global search capability and overcome premature convergence in large-scale mixed-integer optimization. 
Conventional genetic algorithms usually rely on random evolutionary operators, which may lose high-quality solutions or suffer from insufficient exploration in highly coupled UAV-MEC systems. 

To address this issue, we develop an enhanced genetic algorithm (EGA) with elite-guided evolution, heterogeneous variable-aware genetic operations, diversity recovery and adaptive termination strategies. 
The proposed EGA performs global exploration while maintaining population quality and diversity throughout the evolutionary process.

\subsubsection{Elite-guided evolutionary search}
Each individual is encoded as\vspace{-1ex}
\[
\mathcal{X}=(\mathbf{Q},\mathbf{W},\mathbf{f}),
\]
where $\mathbf{Q}$, $\mathbf{W}$ and $\mathbf{f}$ denote UAV trajectories,task offloading decisions and computational resource allocation,respectively.

At each generation, elite individuals are preserved to retain high-quality search directions, while offspring are generated through hybrid crossover and mutation operations. 
Continuous variables, including UAV trajectories and computational resource allocation, adopt convex-combination crossover, whereas discrete offloading decisions use user-level inheritance.

\subsubsection{Diversity-aware evolution control}
To avoid premature convergence caused by excessive exploitation, EGA monitors population diversity and evolutionary stagnation. 
When insufficient diversity is detected, diversity recovery operations are activated to expand the search region. 
Meanwhile, an intelligent early stopping strategy avoids unnecessary iterations when further improvement becomes limited.

By combining elite exploitation and diversity preservation, the adaptive evolution function achieves an effective balance between global exploration and local exploitation.

\subsection{Precision Refinement Function}
\label{sec:prf}

Although EGA provides strong global exploration capability, the obtained solutions may still suffer from limited local accuracy. 
Therefore, HIEA introduces a precision refinement function to further enhance elite solutions through an SCA-based local optimization mechanism.

Different from directly applying conventional SCA, the proposed refinement function is designed for the heterogeneous decision variables of UAV-MEC systems and integrates project-and-convexify smoothing, augmented-objective neighborhood search and adaptive refinement depth control.

\subsubsection{Elite solution refinement}
After each evolutionary generation, the best individual obtained by EGA is selected for refinement. 
The UAV trajectories, task offloading decisions and computational resource allocation are jointly refined through successive convex approximation.

\subsubsection{Project-and-convexify optimization}
To improve local convergence stability, a smoothing operation is introduced:
\[
\tilde{\mathbf{Q}}
=
\mathbf{Q}^{(r)}
+\beta(\mathbf{Q}^{(r)}
-\mathbf{Q}^{(r-1)}),
\]
and similar refinement is applied to other decision variables.

Meanwhile, an augmented objective function is adopted:
\[
\tilde{F}(\mathcal{X})
=
F(\mathcal{X})-\mu V(\mathcal{X}),
\]
where $V(\mathcal{X})$ represents constraint violations.

\subsubsection{Accuracy-efficiency adaptive control}
The refinement depth is controlled by $N_{\mathrm{SCA}}$. 
Increasing the refinement iterations can improve solution accuracy, but introduces additional computational overhead. 
Therefore, HIEA selects a moderate refinement depth to achieve a favorable trade-off between optimization accuracy and computational efficiency.

\section{Theoretical Analysis of HIEA}
\label{sec:theory}

HIEA decomposes the mixed-integer non-convex optimization problem into four coordinated functions: intelligent decision generation, communication-aware initialization, adaptive evolution and precision refinement. 
This section analyzes the key mechanisms of HIEA from three perspectives: search space reduction, initialization quality--diversity balance and accuracy--efficiency trade-off. 
The theoretical observations are further validated through mechanism-oriented 
simulations in Sec.~\ref{sec:mechanism_validation_en}.

\subsection{Search Space Reduction Analysis of MH-DQN}
\label{sec:mhdqn_analysis}

The joint optimization of UAV trajectories, task offloading decisions,
and computational resources leads to a large mixed-integer search space.
Let the original feasible solution space be
\begin{equation}
\Omega=
\Omega_Q\times\Omega_W\times\Omega_f 
\end{equation}
{where $\Omega_Q$, $\Omega_W$ and $\Omega_f$ denote the feasible sets of UAV trajectories, task offloading decisions and computing resource allocation, respectively.}
Instead of directly searching the entire space, MH-DQN learns
priority-aware decision policies and generates promising
trajectory--offloading candidates.
The candidate region is therefore constrained as
\begin{equation}
\Omega_{\mathrm{MH-DQN}}\subseteq\Omega .
\end{equation}

Consequently, the evolutionary optimizer explores a more concentrated
solution region rather than performing blind search over the whole space,
which reduces inefficient exploration and improves convergence efficiency.

\subsection{CQI Guidance Strength and Initialization Quality--Diversity Trade-off}
\label{sec:cqi_analysis}

Although MH-DQN provides high-quality candidate solutions, evolutionary
optimization still requires sufficient population diversity to avoid
premature convergence. Therefore, CQI-LATI introduces communication-aware
trajectory guidance to balance initialization quality and population
diversity.

The trajectory increment is formulated as
\begin{equation}
\Delta\mathbf q_u(t)
=
(1-\alpha_u)
\Delta\mathbf q_{\rm rand}
+
\alpha_u
\Delta\mathbf q_{\rm CQI},
\end{equation}
where $\alpha_u$ denotes the CQI guidance intensity. 
For {ease of analysis}, $\alpha_u$ is {assumed} to be  a constant
$\alpha$.

The initialization quality is measured by
\begin{equation}
\overline{\mathrm{IQ}}(\alpha)
=
\frac{1}{N_pTU}
\sum_{i,t,u}
\frac{1}
{1+
\|\mathbf q_{i,u}(t,\alpha)
-\mathbf q_u^{\mathrm{CQI}}\|_2},
\end{equation}
where $\overline{\mathrm{IQ}}(\alpha)$ characterizes the proximity between
initialized trajectories and communication-efficient service points.

Meanwhile, the population diversity is quantified as
\begin{equation}
\mathrm{Div}_{Q}(\mathcal P_\alpha)
=
\frac{1}{N_p}
\sum_{i=1}^{N_p}
\left\|
\mathbf Q_i(\alpha)
-
\bar{\mathbf Q}(\alpha)
\right\|_2 ,
\end{equation}
where $\mathbf Q_i(\alpha)$ denotes the trajectory component of the
$i$-th initialized individual and $\bar{\mathbf Q}(\alpha)$ represents the
average trajectory of the population.

According to the trajectory update mechanism, increasing $\alpha$ enhances
the contribution of CQI-oriented guidance while reducing random exploration.
Therefore, UAV trajectories become closer to communication-efficient
service points, leading to higher initialization quality. However, the
reduced random perturbation decreases the dispersion among individuals and
may reduce population diversity. This indicates an inherent
quality--diversity trade-off introduced by communication-aware guidance.

To balance exploitation and exploration, CQI-LATI adopts a load-aware
guidance coefficient $\alpha_u$, where heavily loaded clusters receive
stronger communication guidance, while lightly loaded clusters maintain
larger exploration capability. This adaptive strategy improves the
initialization quality without sacrificing population diversity.

\subsection{Accuracy--Efficiency Trade-off of SCA Refinement}
\label{sec:sca_analysis}

After global evolutionary exploration, the obtained solutions may still
suffer from limited local precision. Therefore, HIEA 
introduces an
SCA-based refinement procedure to improve elite solutions.

Let $F^{\mathrm{EGA}}$ denote the best fitness obtained by EGA, where the
fitness represents the objective function value of the optimization problem.
The initial fitness before SCA refinement is given by
\begin{equation}
F(0)=F^{\mathrm{EGA}},
\end{equation}
and $F(n)$ denotes the fitness after $n$ SCA refinement iterations.
With the improvement-based acceptance strategy, only solutions satisfying
\begin{equation}
F(n)\geq F(n-1)
\end{equation}
are retained. Therefore, the refinement process can continuously improve
or maintain the elite solution quality.

However, additional SCA iterations introduce extra optimization and
evaluation overhead. The total runtime can be approximated as
%
\begin{equation}
\mathcal T_{\mathrm{run}}
=
\mathcal T_{\mathrm{EGA}}
+
G_{\mathrm{EGA}}N_{\mathrm{SCA}}\tau_{\mathrm{SCA}}
+
\mathcal T_{\mathrm{oth}}
\end{equation}
where $G_{\mathrm{EGA}}$ denotes the number of EGA generations,
$\tau_{\mathrm{SCA}}$ denotes the cost of one SCA iteration, and
$\mathcal T_{\mathrm{oth}}$ represents other computational overheads.

Hence, increasing $N_{\mathrm{SCA}}$ improves local refinement accuracy
but increases computational cost. When the solution improvement becomes
saturated, further refinement provides limited benefits, resulting in an
accuracy--efficiency trade-off.

\subsection{Unified Design Implications}

The above analysis reveals three fundamental mechanisms of HIEA:
\begin{align}
|\Omega_{\mathrm{MH-DQN}}|
&\ll |\Omega|,
&&\text{(Search Space Reduction)}
\\
\overline{\mathrm{IQ}}
&\leftrightarrow
\mathrm{Div}(\mathcal P),
&&\text{(Initialization Balance)}
\\
F
&\leftrightarrow
\mathcal T_{\mathrm{run}},
&&\text{(Refinement Efficiency)}
\end{align}

Therefore, HIEA does not rely on a single optimization strategy.
Instead, it achieves hierarchical optimization by integrating policy-guided decision generation, communication-aware initialization, evolutionary global exploration and precision refinement.

\begin{table}[t]
\centering
\caption{Main Simulation Parameters}
\label{tab:sim_params}
\renewcommand{\arraystretch}{1.05}
\setlength{\tabcolsep}{3pt}
\footnotesize

\begin{tabular}{@{}p{0.42\linewidth}p{0.45\linewidth}@{}}
\hline
\textbf{Parameter} & \textbf{Value} \\
\hline

Number of UAVs/users (default)
& $U=4, K=60$ \\

Service area
& $500\times500\,\mathrm{m}^{2}$ \\

UAV altitude
& $z_u(t)\in[20,100]\,\mathrm{m}$ \\

Maximum UAV velocity
& $V_{\max}=8\,\mathrm{m/s}$ \\

Carrier frequency
& $f_c=5.8\,\mathrm{GHz}$ \\

Bandwidth
& $B=20\,\mathrm{MHz}$ \\

Noise power spectral density
& $N_0=-174\,\mathrm{dBm/Hz}$ \\

CPU frequency
& $f_k^{\mathrm{loc}}=2\,\mathrm{GHz},~
f_u=10\,\mathrm{GHz}$ \\

Task data size
& $n_k\in[0.5,5]\,\mathrm{MB}$ \\

CPU cycles per bit
& $c_k\in[500,2000]$ cycles/bit \\

High/low priority ratio
& $40\%/60\%$ \\

\hline
\end{tabular}
\end{table}

\begin{figure*}[t]
\centering
\small  

\begin{minipage}{0.31\textwidth}
  \centering
  \includegraphics[trim=0mm 7mm 5mm 0mm, width=\linewidth]{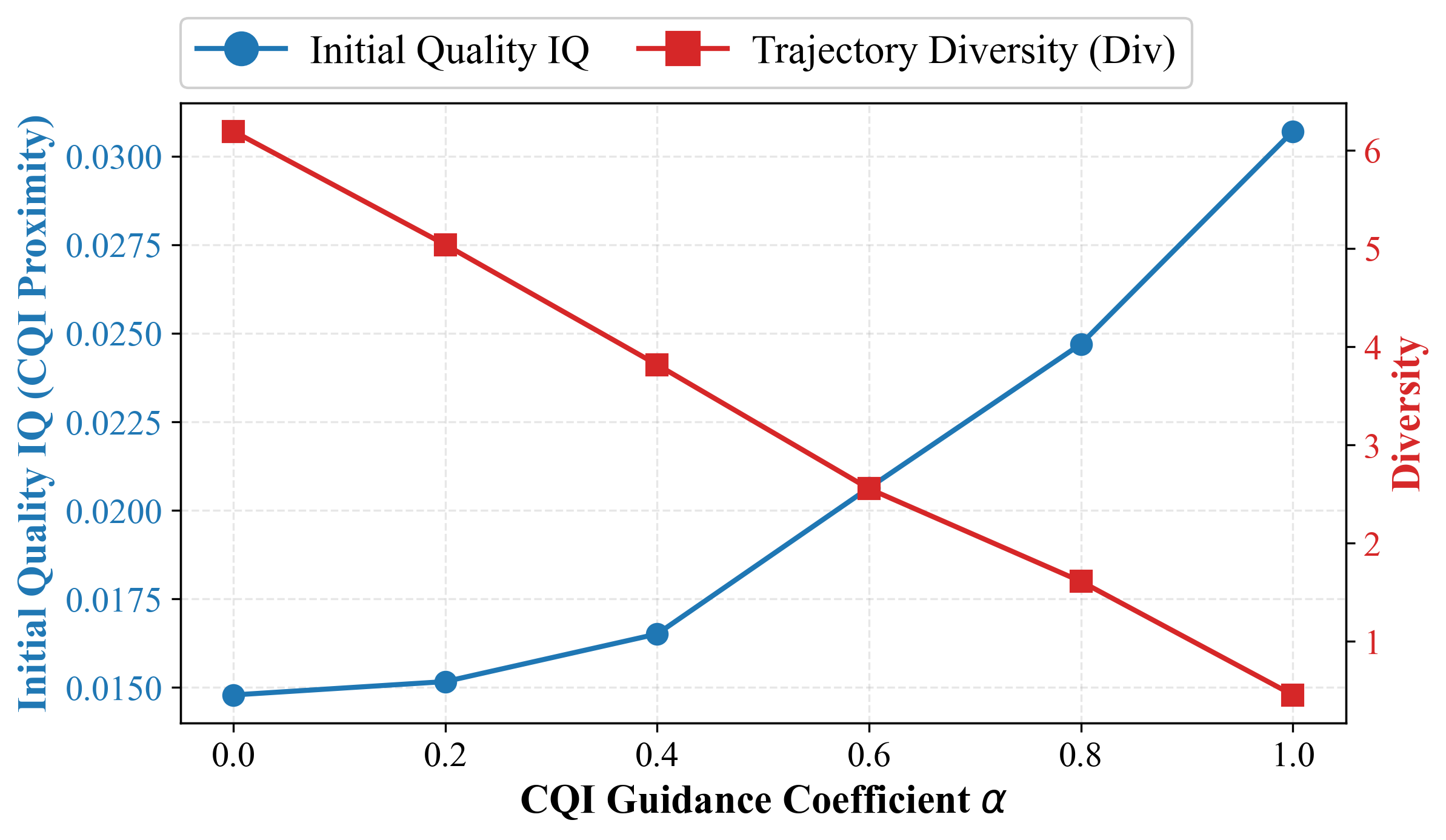}
  \caption{CQI guidance coefficient $\alpha$ and the IQ--Div trade-off}
  \label{fig:alpha_tradeoff}
\end{minipage}\hfill
\begin{minipage}{0.31\textwidth}
  \centering
  \includegraphics[trim=5mm 5mm 5mm 0mm, width=\linewidth]{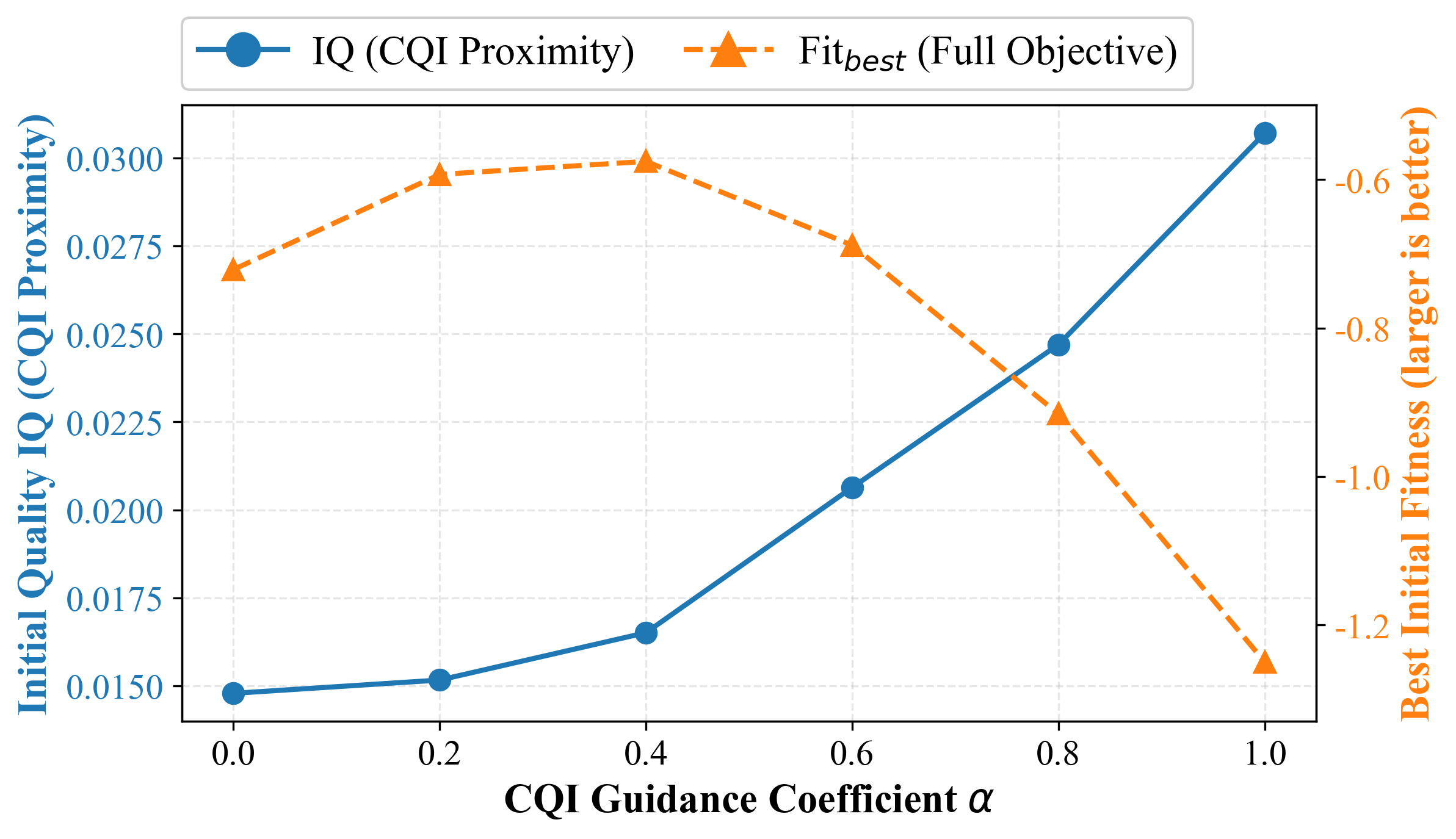}
  \caption{Initialization quality--fitness trade-off vs.  CQI guidance intensities}
  \label{fig:iq_fitbest}
\end{minipage}\hfill
\begin{minipage}{0.31\textwidth}
  \centering
  \includegraphics[trim=8mm 4mm -8mm 1mm, width=1.08\linewidth]{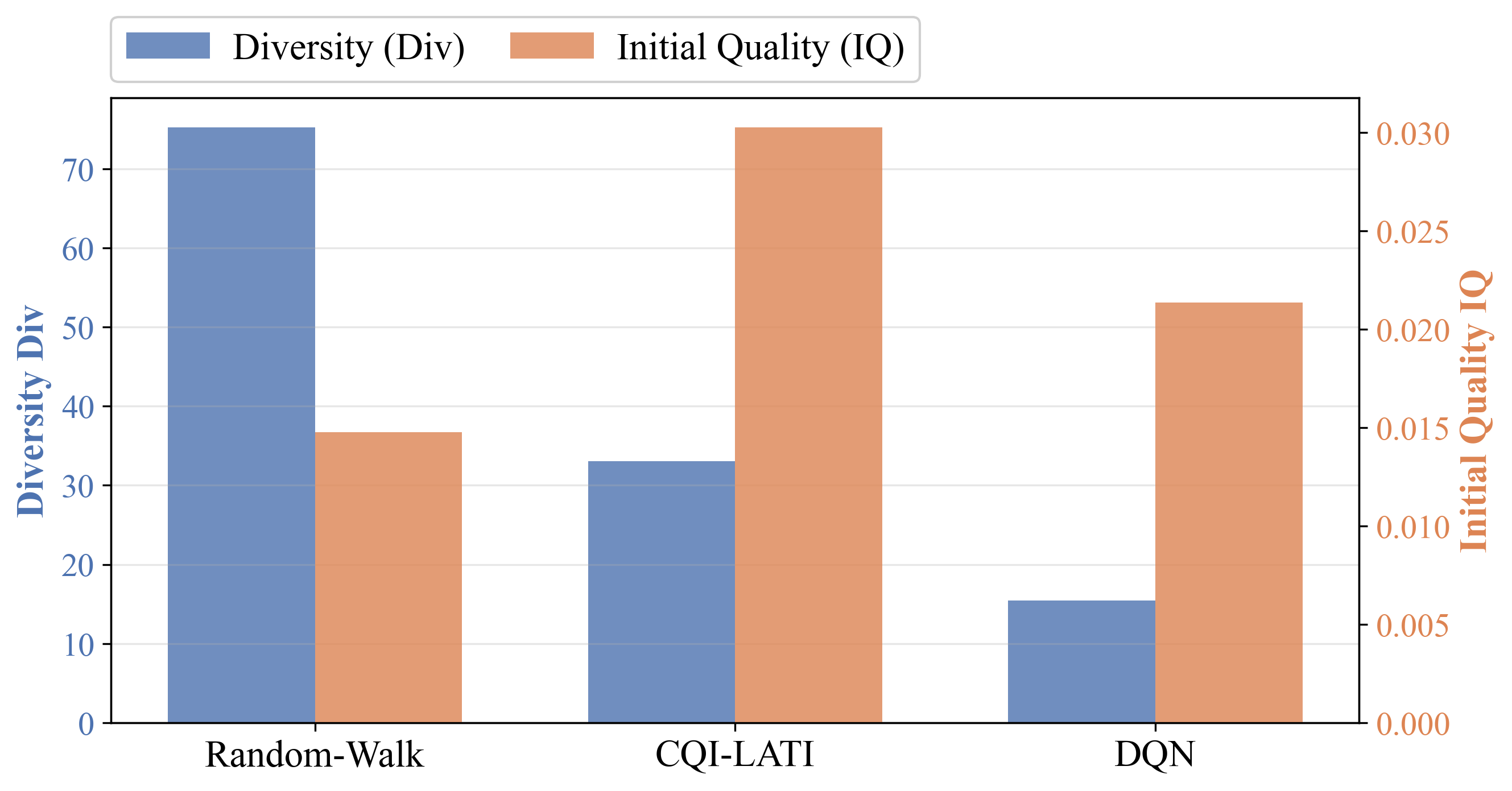}
  \caption{Quality--diversity comparison of different initialization strategies}
  \label{fig:init_tradeoff}
\end{minipage}

\vspace{2pt}

\begin{minipage}{0.31\textwidth}
  \centering
  \includegraphics[trim=5mm 5mm 0mm -5mm, width=1.1\linewidth]{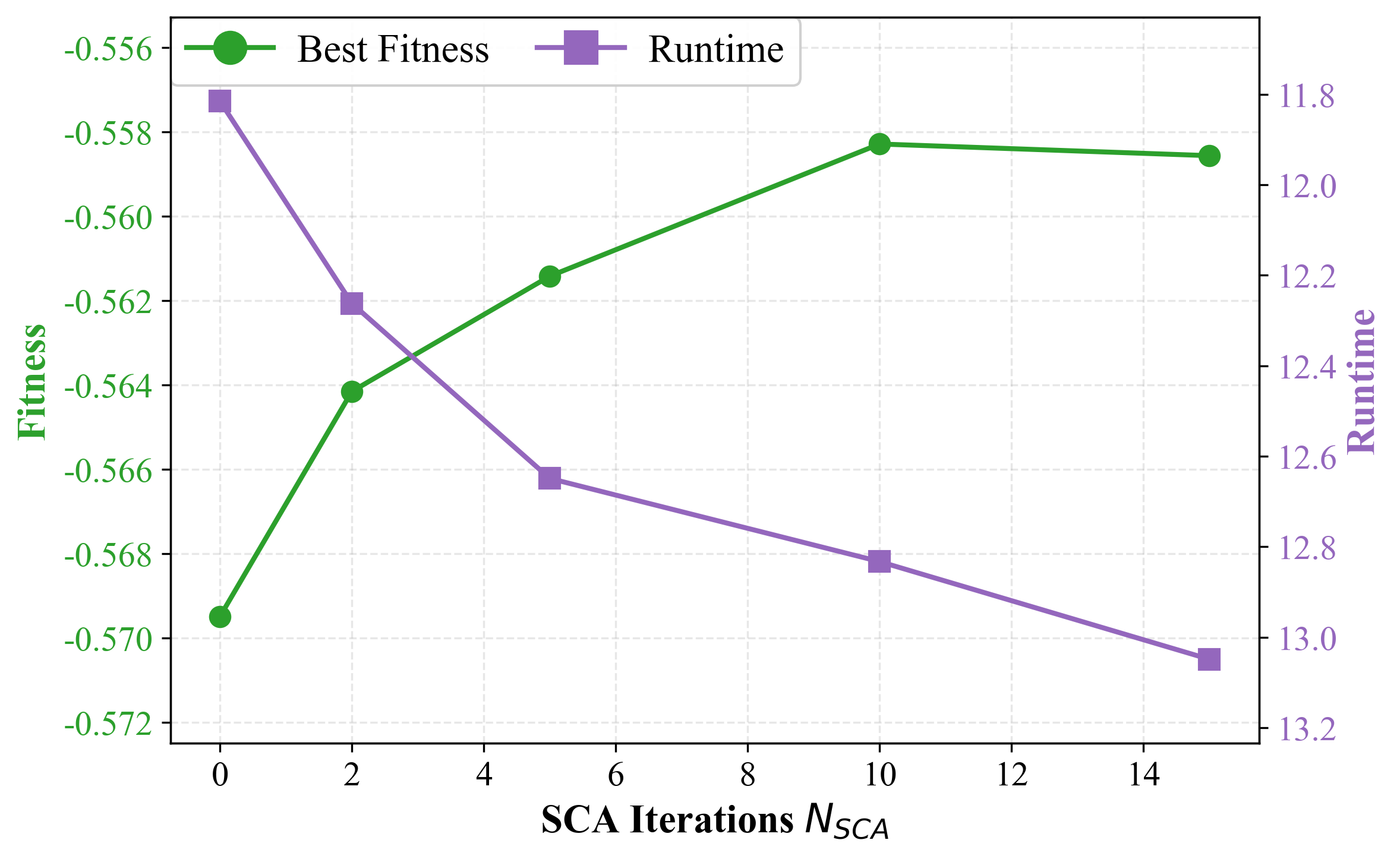}
  \caption{SCA Accuracy-efficiency trade-off 
  under different refinement depths}
  \label{fig:sca_tradeoff}
\end{minipage}\hfill
\begin{minipage}{0.31\textwidth}
  \centering
  \includegraphics[trim=0mm 4mm 0mm 7.5mm, width=1.1\linewidth]{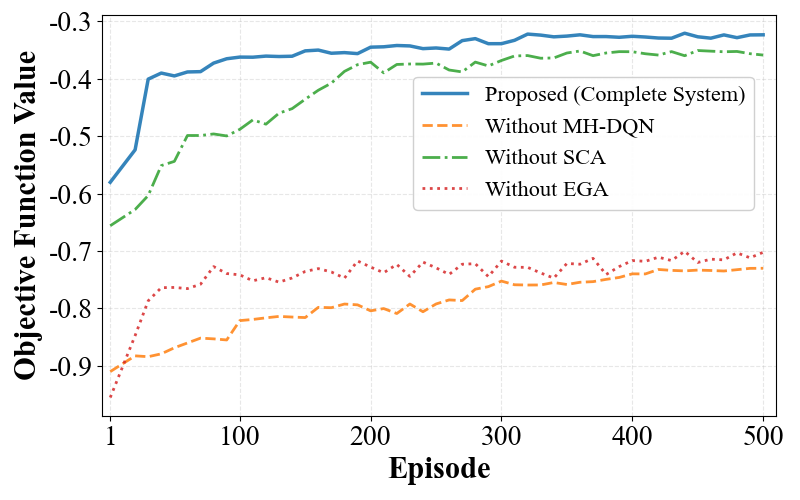}
  \caption{Module ablation}
  \label{fig:ablation}
\end{minipage}\hfill
\begin{minipage}{0.31\textwidth}
  \centering
  \includegraphics[trim=0mm 4mm 4mm -4mm, width=0.95\linewidth]{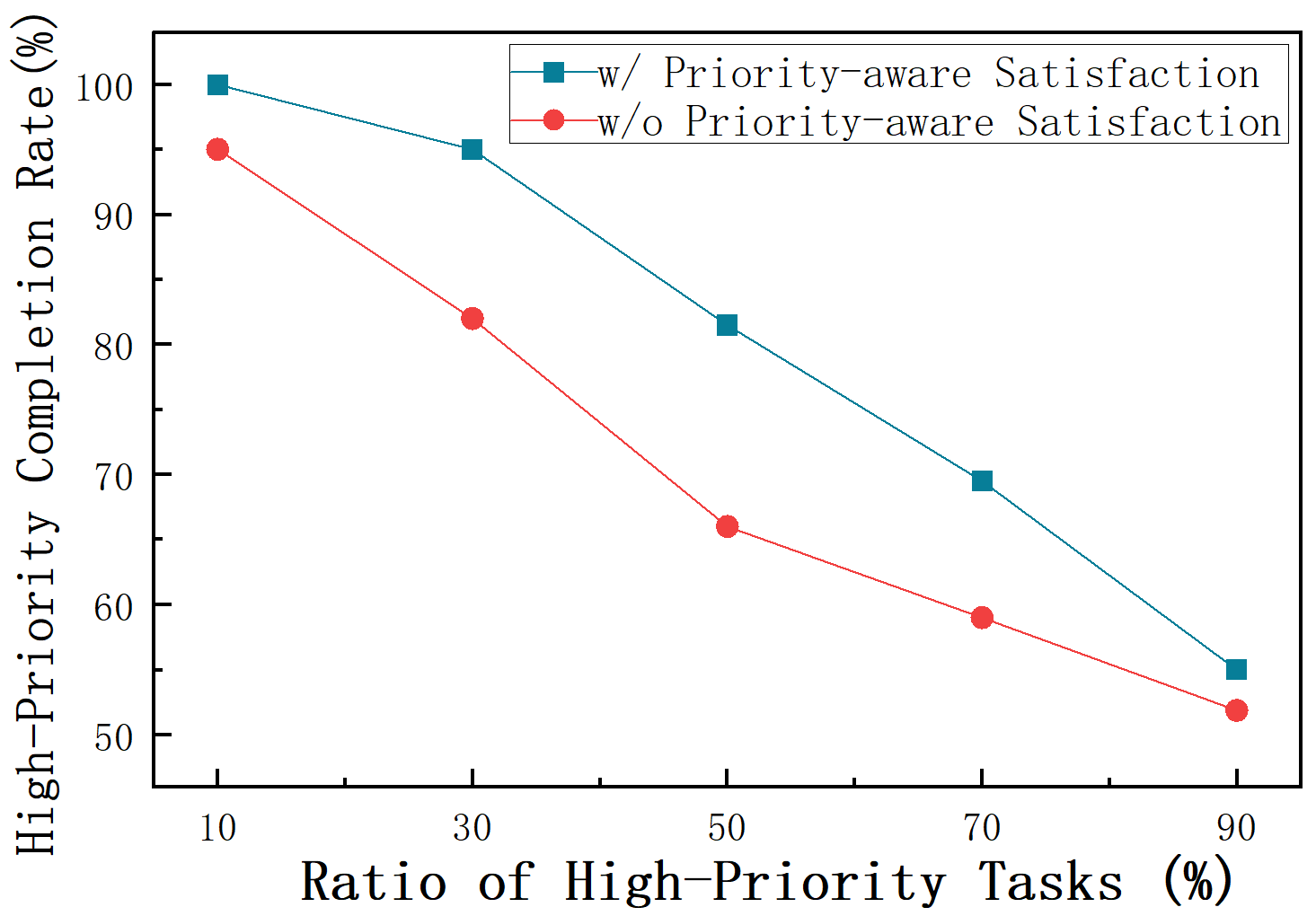}
  \caption{Task completion rate under different high-priority ratios}
  \label{fig:priority}
\end{minipage}
\end{figure*}

\section{SIMULATION RESULTS AND DISCUSSION}
\label{sec:simulation_setup_en}

This section evaluates HIEA in a UAV-assisted MEC network.
The main simulation parameters are summarized in Table~\ref{tab:sim_params}.
Each set of simulations is repeated with multiple random seeds.

Unless otherwise specified, the default network configuration is adopted,
where $4$ UAVs serve $60$ mobile users.
Users are associated with UAV clusters via 
a K-means-based user clustering method 
with periodic re-clustering and uplink NOMA is employed within each cluster.
The proposed optimizer jointly determines UAV trajectories, task offloading
decisions and computing resources under the considered throughput-energy objective.

For mechanism validation, only the corresponding factor, including the CQI
guidance coefficient $\alpha$, SCA refinement depth $N_{\mathrm{SCA}}$,
or initialization strategy, is varied while other system settings remain
unchanged. 

\subsection{Mechanism Validation}
\label{sec:mechanism_validation_en}

The theoretical analysis in Sec.~\ref{sec:theory} reveals two key
trade-offs in HIEA, i.e., initialization quality--diversity balance and
refinement accuracy--efficiency trade-off.
This subsection validates these mechanisms through controlled simulations. 

\subsubsection{Validation of CQI-LATI Initialization}

The proposed CQI-LATI balances communication guidance and population
diversity through the adaptive coefficient $\alpha$.
As shown in Fig.~\ref{fig:alpha_tradeoff}, increasing $\alpha$ improves
the initialization quality $\overline{\mathrm{IQ}}$ but reduces trajectory
diversity, confirming the inherent quality--diversity trade-off.

Moreover, Fig.~\ref{fig:iq_fitbest} shows that excessively strong CQI
guidance does not always improve the final optimization performance.
Therefore, the load-aware guidance strategy achieves a better balance
between communication exploitation and evolutionary exploration.

Fig.~\ref{fig:init_tradeoff} further compares different initialization
strategies. CQI-LATI achieves higher initialization quality than random
initialization while preserving more diversity than 
{DQN-based initialization}, demonstrating its effectiveness as an internal function
of HIEA.

\subsubsection{Validation of SCA Refinement}

Fig.~\ref{fig:sca_tradeoff} evaluates the impact of refinement depth
$N_{\mathrm{SCA}}$. Increasing $N_{\mathrm{SCA}}$ improves solution quality
but also increases computational cost, with diminishing performance gains
in later iterations. This verifies the accuracy--efficiency trade-off of
SCA refinement.

Overall, the results confirm that HIEA benefits from the coordinated
design of communication-aware initialization and precision refinement.

\subsection{Ablation Study}

Fig.~\ref{fig:ablation} evaluates the contribution of
the MH-DQN, EGA and SCA modules. The complete HIEA
framework consistently achieves the highest objective value,
demonstrating the complementary benefits of the three
components.

{Fig.~\ref{fig:priority} evaluates the priority-aware satisfaction model under different high-priority task ratios.
As the ratio increases, the task completion rate decreases due to resource competition.
The proposed model consistently outperforms the scheme without priority awareness, demonstrating its effectiveness in protecting high-priority task services.}

\begin{figure}[b]
\centering    
\includegraphics[trim=0mm 0mm 0mm 5mm, width=0.75\columnwidth]{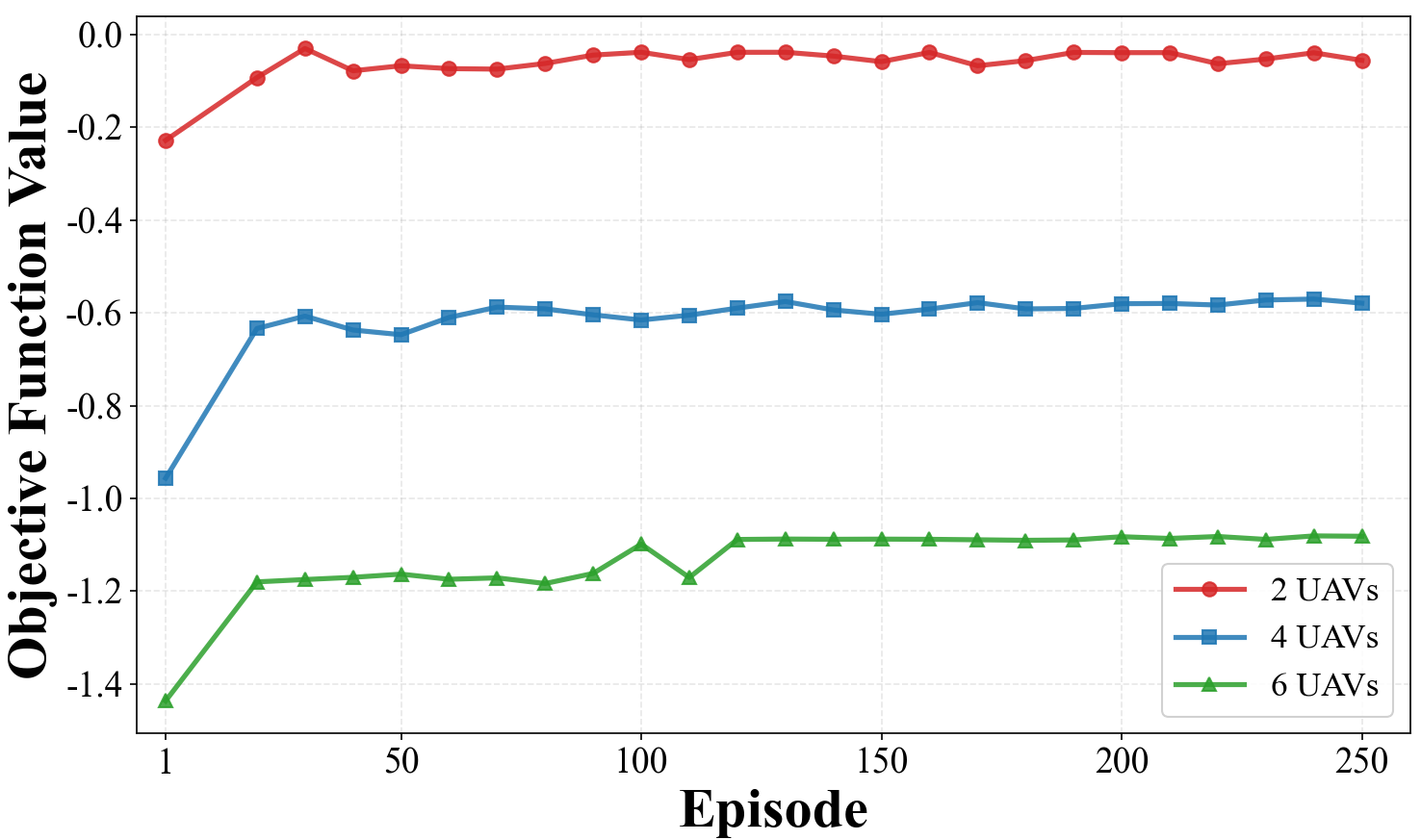}    
\caption{Convergence Performance of the Proposed Algorithm under Different Numbers of UAVs}    
\label{fig:Convergence}
\end{figure}

\begin{figure*}[t]
\centering
\small  

\begin{minipage}{0.31\textwidth}
  \centering
  \includegraphics[trim=0mm 7mm 0mm 0mm, width=\linewidth]{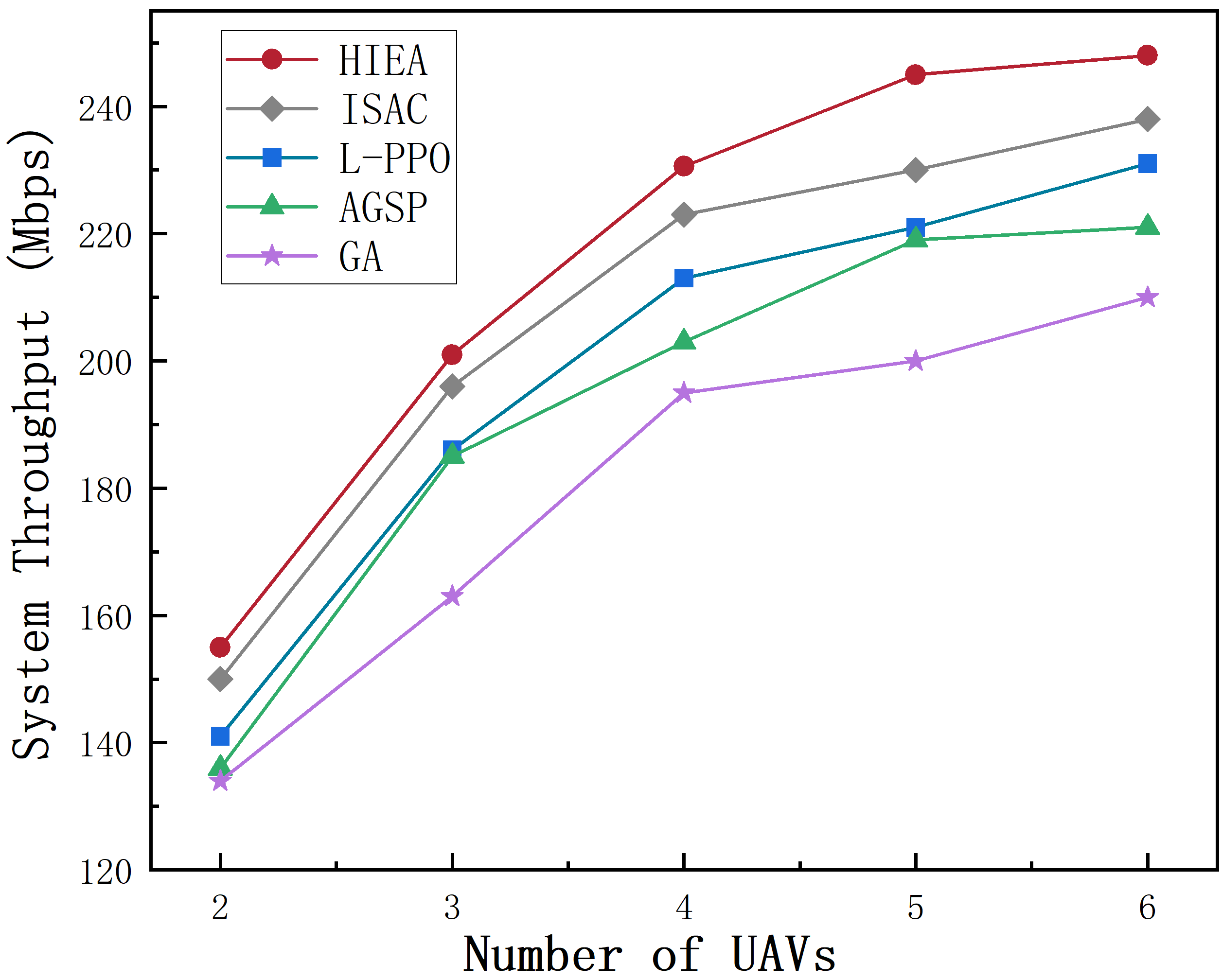}
  \caption{System throughput versus the number of UAVs}
  \label{fig:throughput_ue}
\end{minipage}\hfill
\begin{minipage}{0.31\textwidth}
  \centering
  \includegraphics[trim=0mm 5mm 0mm 0mm, width=0.975\linewidth]{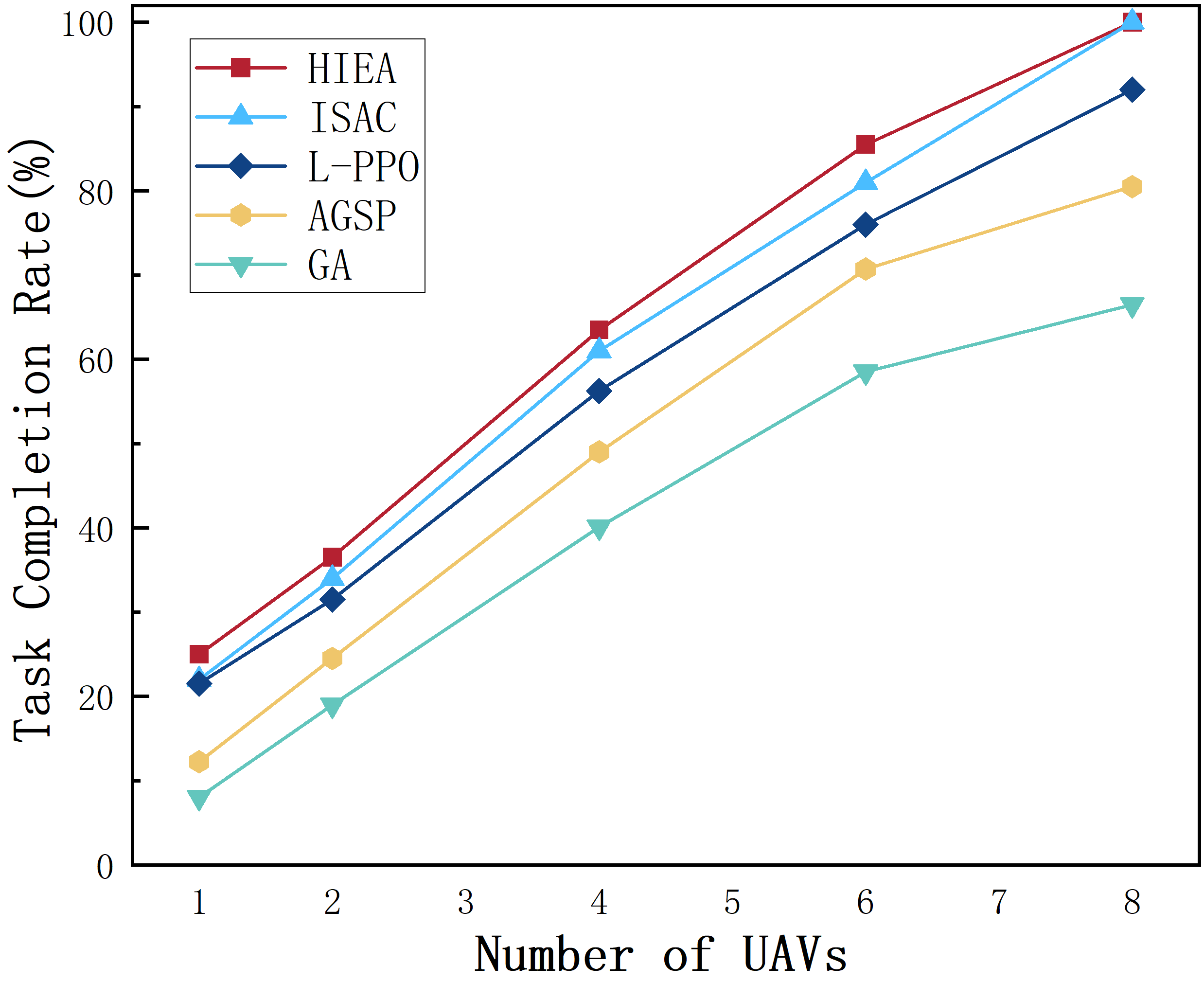}
  \caption{Task completion ratio versus the number of UAVs}
  \label{fig:completion_uav}
\end{minipage}\hfill
\begin{minipage}{0.31\textwidth}
  \centering
  \includegraphics[trim=0mm 4mm 0mm 1mm, width=\linewidth]{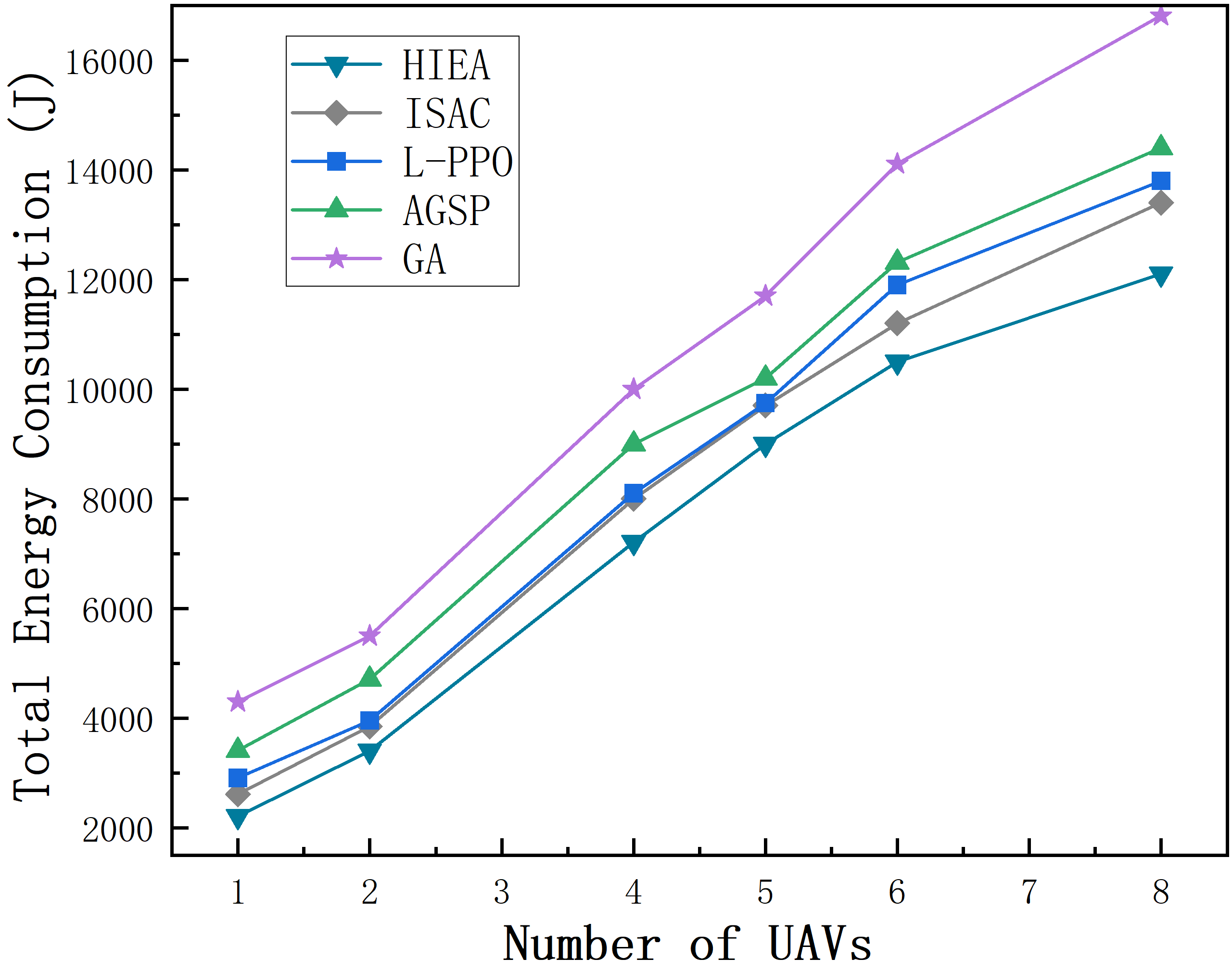}
  \caption{Total system energy consumption versus the number of UAVs}
  \label{fig:energy_uav}
\end{minipage}

\vspace{2pt}

\begin{minipage}{0.31\textwidth}
  \centering
  \includegraphics[trim=0mm 5mm 0mm -5mm, width=\linewidth]{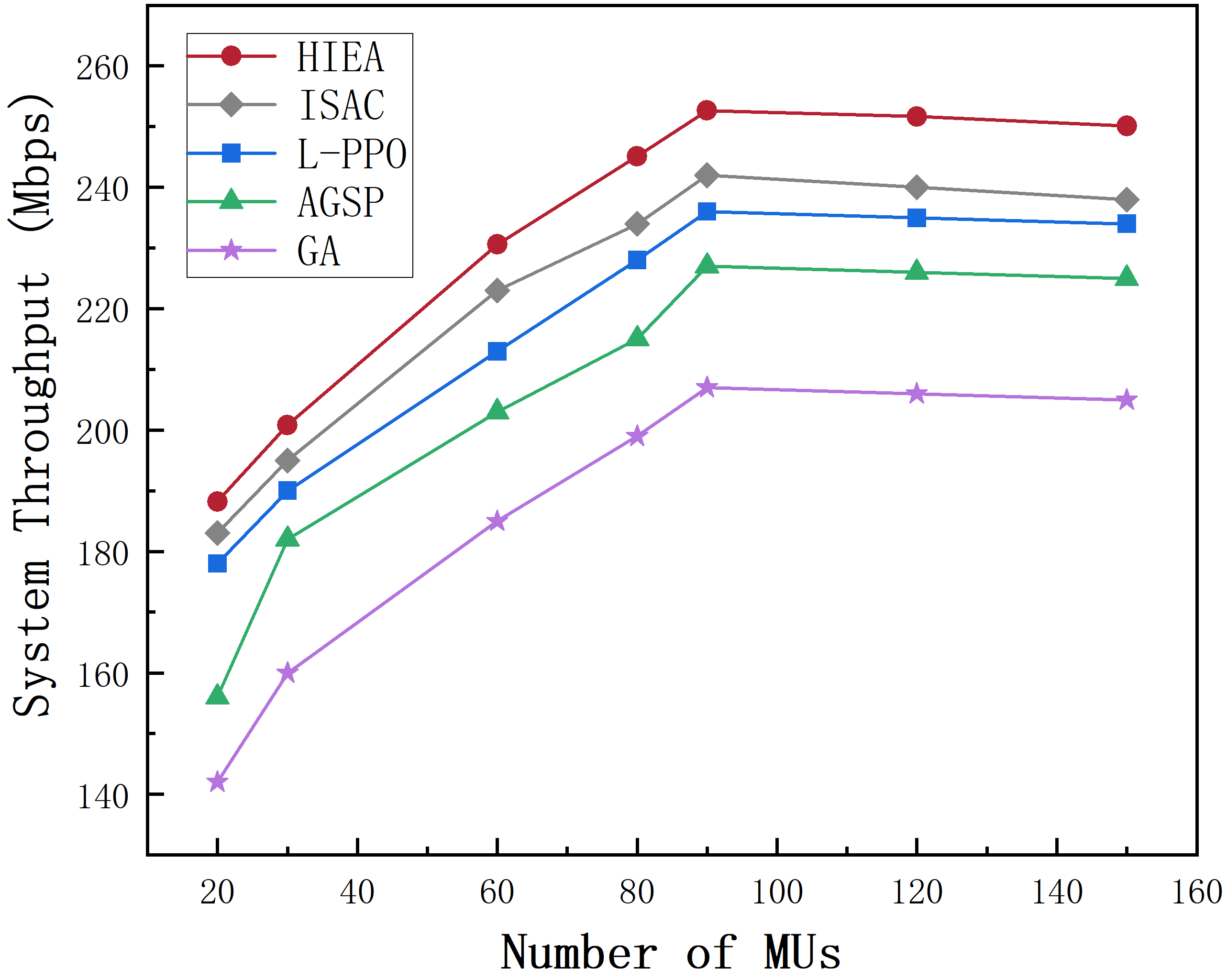}
  \caption{System throughput versus the number of MUs}
  \label{fig:throughput_uav}
\end{minipage}\hfill
\begin{minipage}{0.31\textwidth}
  \centering
  \includegraphics[trim=0mm 4.5mm 0mm -5mm, width=0.975\linewidth]{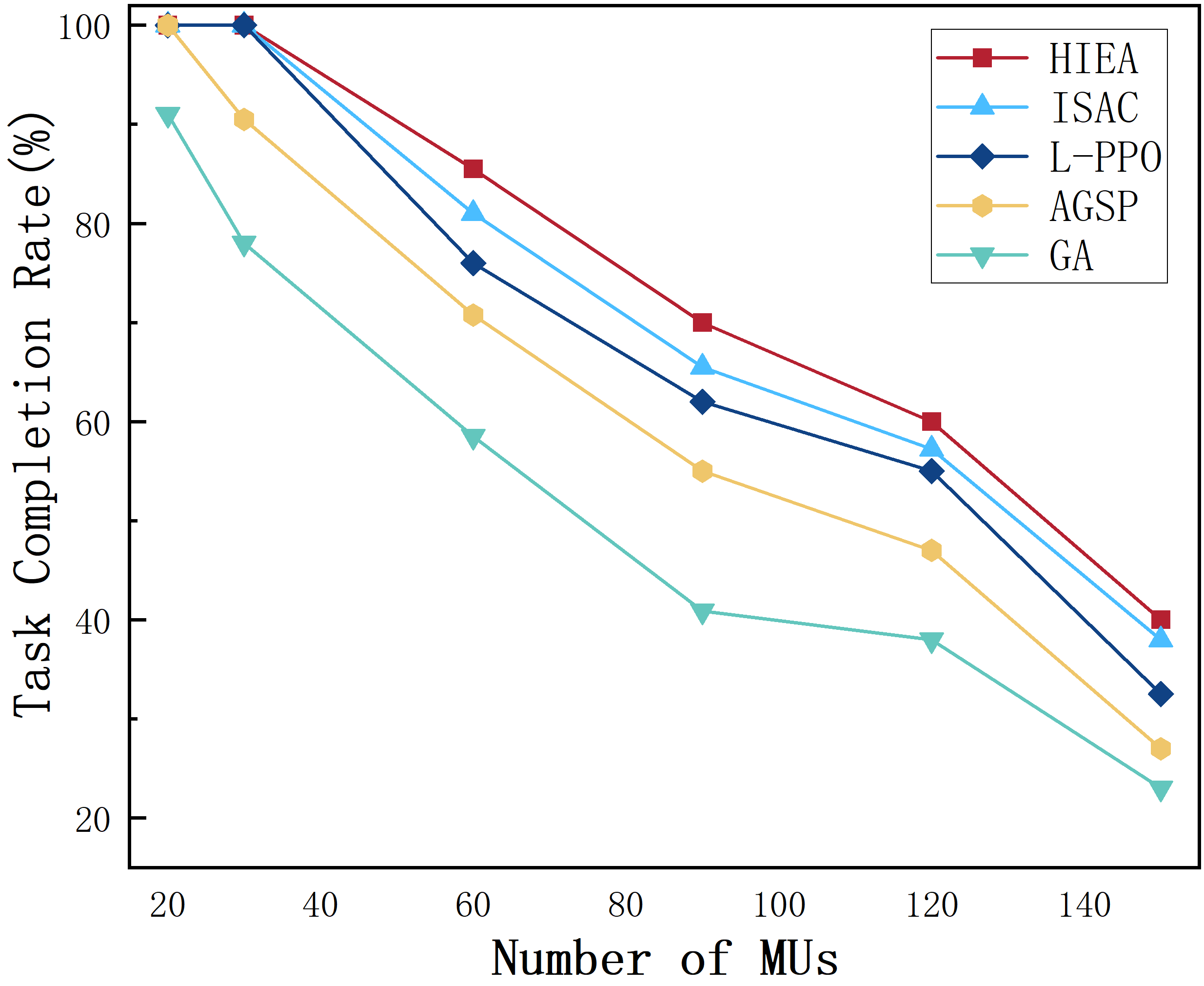}
  \caption{Task completion ratio versus the number of MUs}
  \label{fig:completion_ue}
\end{minipage}\hfill
\begin{minipage}{0.31\textwidth}
  \centering
  \includegraphics[trim=0mm 4mm 0mm -4mm, width=\linewidth]{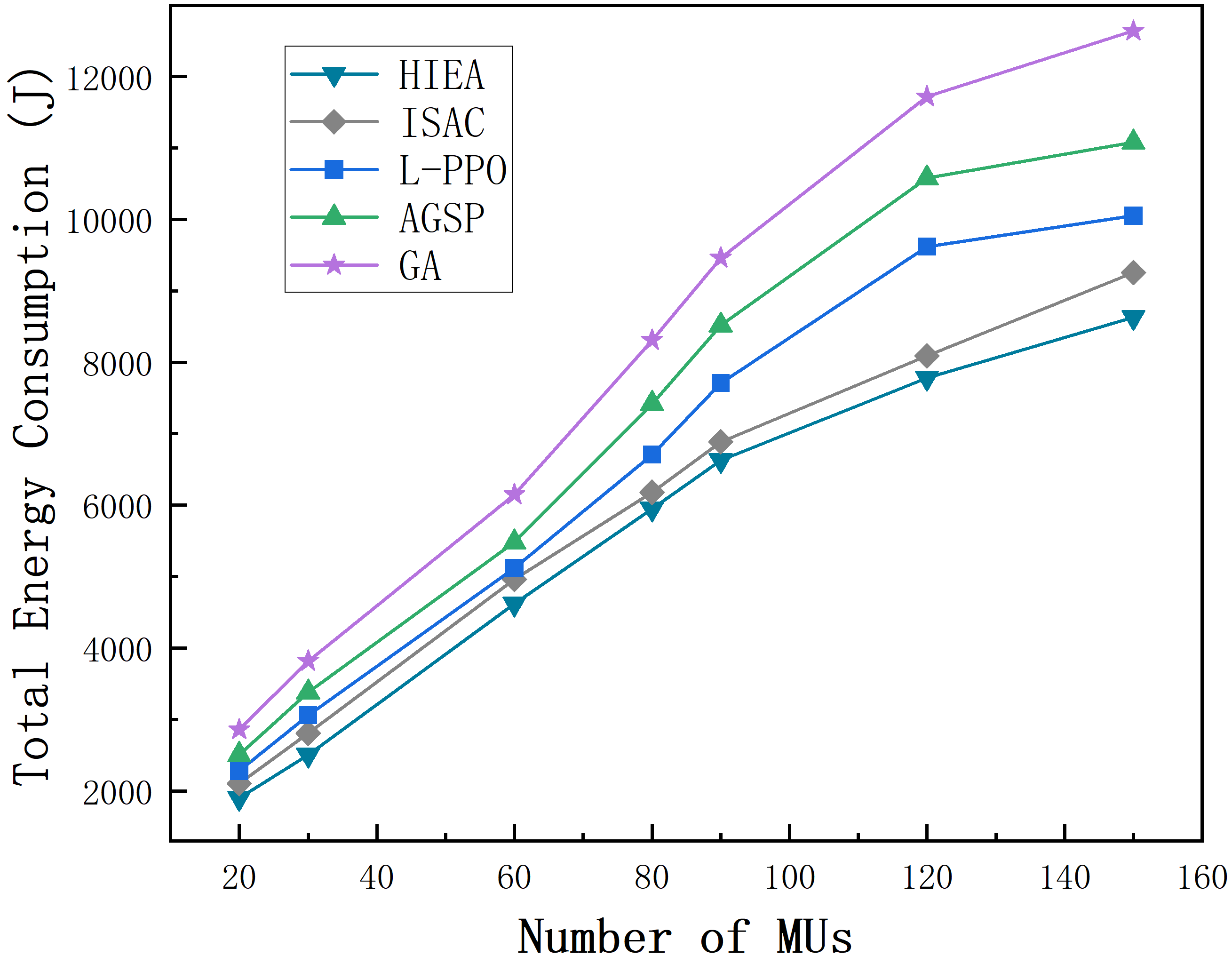}
  \caption{Total system energy consumption versus the number of MUs}
  \label{fig:energy_ue}
\end{minipage}

\end{figure*}

\subsection{Convergence Analysis}
Fig.~\ref{fig:Convergence} shows the convergence performance under different numbers of UAVs. All curves increase rapidly in the early episodes and then stabilize, indicating fast convergence of the proposed algorithm. The objective function becomes stable after about 100 episodes, with only slight fluctuations.

It can also be observed that the performance decreases slightly as the number of UAVs increases, due to higher system complexity and resource coupling. However, all cases achieve stable convergence, demonstrating the robustness of the proposed method.

Moreover, the SCA-based refinement introduces limited overhead by only
optimizing elite solutions, achieving a favorable solution-quality versus
complexity trade-off.

\subsection{Baseline Comparison}
In this section, 
HIEA is evaluated against several benchmark schemes, including AGSP~\cite{yuan2024cost}, Genetic Algorithm (GA), ISAC~\cite{chen2025qos} and PPO-L~\cite{he2025joint}. AGSP combines GA, PSO and SA to enhance optimization performance, while GA uses standard evolutionary operations. ISAC is based on the Soft Actor-Critic framework with a perturbation mechanism for improved exploration. PPO-L integrates Louvain clustering for user grouping, a load penalty for balancing resources and Proximal Policy Optimization for task scheduling.

The performance is evaluated using three metrics: system throughput (Mbps), task completion ratio (\%) and total energy consumption (J).

Fig.~\ref{fig:throughput_ue} and Fig.~\ref{fig:throughput_uav} show the system throughput under different numbers of UAVs and MUs, respectively.
As the number of MUs increases, all algorithms improve and gradually saturate due to limited resources.
Increasing the number of UAVs further improves throughput, while the gain gradually diminishes.
HIEA achieves the highest throughput in both scenarios, 
improving the throughput over ISAC by up to 5.1\% and 6.5\%, respectively.

Fig.~\ref{fig:completion_uav} and Fig.~\ref{fig:completion_ue} compare the task completion ratio under scenarios of varying 
numbers of UAVs and MUs, respectively.
Increasing the number of UAVs improves the completion ratio by providing additional communication and computing resources, while increasing the number of MUs causes performance degradation due to resource competition.
HIEA achieves the highest completion ratio in both scenarios, 
improving the task completion ratio over ISAC by up to 13.6\% and 6.9\% in the UAV and MU scenarios, 
respectively.

Lastly, we compare the total system energy consumption. Fig.~\ref{fig:energy_uav} and Fig.~\ref{fig:energy_ue} illustrate the total system energy consumption under different numbers of UAVs and MUs, respectively.
The energy consumption increases with the numbers of UAVs and MUs due to additional communication, computation, and operating overhead.
HIEA achieves the lowest energy consumption among all methods, reducing energy consumption over ISAC by up to 15.3\% in the UAV scenario and up to 10.7\% in the MU scenario. 

\section{Conclusion}

This paper investigated the priority-aware multi-UAV cooperative NOMA-MEC optimization problem and proposed a Hierarchical Intelligent Evolutionary Algorithm (HIEA) to improve optimization efficiency through intelligent decision guidance, communication-aware initialization, evolutionary search and refinement. 
Through theoretical analysis, we characterized HIEA’s search-space reduction, the
quality-diversity trade-off of CQI-guided initialization, and the
accuracy-efficiency trade-off of SCA refinement. 
Simulation results then verified that HIEA outperforms benchmark methods in terms of throughput, task completion rate and energy efficiency. 
Future work will focus on developing dedicated optimization algorithms for high-priority UAVs to further improve emergency task services in dynamic multi-UAV networks.

\nocite{*}
\bibliographystyle{IEEEtran}
\balance
\small
\bibliography{reference}

\end{document}